\documentclass[sigconf]{acmart}

\usepackage{amsmath,amsfonts} 
\usepackage{lipsum} 
\usepackage{algorithm,balance,pbox}
\usepackage{algorithmicx}
\usepackage{algpseudocode}
\usepackage{graphicx}
\usepackage{subfigure}
\usepackage{multirow}
\usepackage{hhline}
\usepackage{array,multirow,diagbox}
\usepackage{color,booktabs}
\usepackage{colortbl}

\newcommand*{\thead}[1]{%
\multicolumn{1}{c}{\bfseries\begin{tabular}{@{}c@{}}#1\end{tabular}}}

\copyrightyear{2019} 
\acmYear{2019} 
\setcopyright{acmcopyright}
\acmConference[WSDM '19]{The Twelfth ACM International Conference on Web Search and Data Mining}{February 11--15, 2019}{Melbourne, VIC, Australia}
\acmBooktitle{The Twelfth ACM International Conference on Web Search and Data Mining (WSDM '19), February 11--15, 2019, Melbourne, VIC, Australia}
\acmPrice{15.00}
\acmDOI{10.1145/3289600.3290994}
\acmISBN{978-1-4503-5940-5/19/02}
\begin{document}
\title[The Role of Social Context for Fake News Detection]{Beyond News Contents:\\The Role of Social Context for Fake News Detection}

\author{Kai Shu}
\affiliation{%
  \institution{Arizona State University}
}
\email{kai.shu@asu.edu}

\author{Suhang Wang}
\affiliation{%
  \institution{Penn State University}
}
\email{szw494@psu.edu}

\author{Huan Liu}
\affiliation{%
  \institution{Arizona State University}
}
\email{huan.liu@asu.edu}

\begin{abstract}
Social media is becoming popular for news consumption due to its fast dissemination, easy access, and low cost. However, it also enables the wide propagation of \textit{fake news}, i.e.,  news with intentionally false information. Detecting fake news is an important task, which not only ensures users receive authentic information but also helps maintain a trustworthy news ecosystem. The majority of existing detection algorithms focus on finding clues from news contents, which are generally not effective because fake news is often intentionally written to mislead users by mimicking true news. Therefore, we need to explore auxiliary information to improve detection.  The social context during news dissemination process on social media forms the inherent \textit{tri-relationship}, the relationship among publishers, news pieces, and users, which has potential to improve fake news detection. For example, partisan-biased publishers are more likely to publish fake news, and low-credible users are more likely to share fake news. In this paper, we study the novel problem of exploiting social context for fake news detection. We propose a tri-relationship embedding framework TriFN, which models publisher-news relations and user-news interactions simultaneously for fake news classification. We conduct experiments on two real-world datasets, which demonstrate that the proposed approach significantly outperforms other baseline methods for fake news detection.
\end{abstract}


\maketitle

\section{Introduction}\label{sec:introduction}
People nowadays tend to seek out and consume news from social media rather than traditional news organizations.  For example, 62\% of U.S. adults get news on social media in 2016, while in 2012, only 49 percent is reported seeing news on social media\footnote{http://www.journalism.org/2016/05/26/news-use-across-social-media-platforms-2016/}.  However, social media is a double-edged sword for news consumption. The quality of news on social media is much lower than that of traditional news organizations. Large volumes of \textit{fake news}, i.e., news with intentionally false information, are produced online for a variety of purposes, such as financial and political gain~\cite{klein2017fake,allcott2017social}.


Fake news can have detrimental effects on individuals and the society. First, people may be misled by fake news and accept false beliefs~\cite{nyhan2010corrections,paul2016russian}.  Second, fake news could change the way people respond to true news\footnote{https://www.nytimes.com/2016/11/28/opinion/fake-news-and-the-internet-shell-game.html?}. Third, the wide propagation of fake news could break the trustworthiness of entire news ecosystem. Thus, it is important to detect fake news on social media. Fake news is intentionally written to mislead consumers, which makes it nontrivial to detect simply based on news content. To build an effective and practical fake news detection system, it is natural and necessary to explore \textit{auxiliary information} from different perspectives.


The news ecosystem on social media provides abundant social context information, which involves three basic entities, i.e., publishers, news pieces, and social media users. Figure~\ref{fig:example} gives an illustration of such ecosystem. In Figure~\ref{fig:example}, $p_1$, $p_2$ and $p_3$ are news publishers who publish news $a_1, \dots, a_4$ and $u_1, \dots, u_6$ are users who have engaged in sharing these news pieces. In addition, users tend to form social links with like-minded people with similar interests. As we will show, the \textit{tri-relationship}, the relationship among publishers, news pieces, and users, contains additional information to help detect fake news. 

First, sociological studies on journalism have theorized the correlation between the partisan bias of publisher and the veracity degree of news content~\cite{gentzkow2014media}. The partisan bias means the perceived bias of the publisher in the selection of how news is reported and covered~\cite{entman2007framing}. For example, in Figure~\ref{fig:example}, $p_1$ is a publisher with extreme left partisan bias and $p_2$ is a publisher with extreme right partisan bias. To support their own partisan viewpoints, they have high degree to distort the facts and report fake news pieces, such as $a_1$ and $a_3$; while for a mainstream publisher $p_3$ that has least partisan bias, he/she has a lower chance to manipulate original news events, and is more likely to write a true news piece $a_4$. Thus, exploiting the partisan bias of publishers to bridge the publisher-news relationships can bring additional benefits to predict fake news. 

Second, mining user engagements towards news pieces on social media also help fake news detection. Previous approaches try to aggregate users' attributes to infer the degree of news veracity by assuming that either (i) all the users contribute equally for learning feature representations of news pieces~\cite{jin2016news}; or (ii) user features are grouped locally for specific news and the global user-news interactions are ignored~\cite{castillo2011information}. However, in practice, these assumptions may not hold. On social media, different users have different credibility levels. The credibility score, which means ``the quality of being trustworthy''~\cite{abbasi2013measuring}, has a strong indication of whether some user is more likely to share fake news or not. Those less credible users, such as malicious accounts or normal users who are vulnerable to fake news, are more likely to spread fake news. For example, $u_2$ and $u_4$ are users with low credibility scores, and they tend to spread fake news more than other highly credible users. In addition, users tend to form relationships with like-minded people~\cite{quattrociocchi2016echo}. For example, user $u_5$ and $u_6$ are friends on social media, so they tend to post those news that confirm their own views, such as $a_4$. Therefore, incorporating the user credibility levels to capture the user-news interactions has potentials to improve fake news prediction.

Moreover, the publisher-news relationships and user-news interactions both provide new and different perspectives of social context, and thus contain complementary information to advance fake news detection.  In this paper, we investigate: (1) how to mathematically model the tri-relationship to extract feature representations of news pieces; and (2) how to take advantage of tri-relationship modeling for fake news detection. Our solutions to these two challenges results in a novel framework TriFN for fake news detection problem. Our main contributions are summarized as follows:

\begin{figure}[tbp!]
   \centering
   \includegraphics[width=0.6\linewidth]{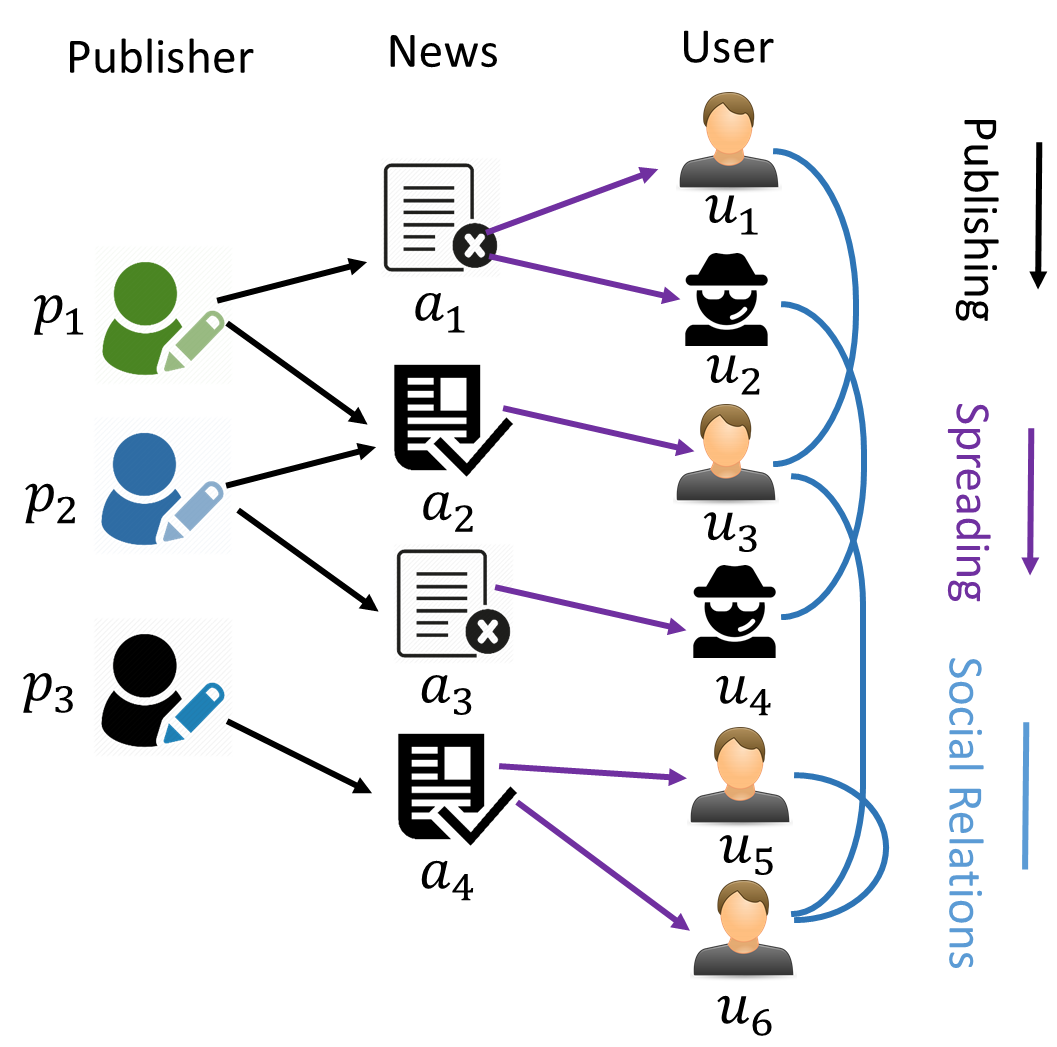}
   \caption{An illustration of tri-relationship among publishers, news pieces, and users, during the news dissemination process. For example, an edge $(p\rightarrow a)$ demonstrates that publisher $p$ publishes news item $a$, an edge $(a\rightarrow u)$ represents news item $a$ is spread by user $u$, and an edge $(u_1 \leftrightarrow u_2)$ indicates the social relation between user $u_1$ and $u_2$. }\vspace{-0.25cm}
   \label{fig:example}
\end{figure}


\begin{itemize}
  \item We provide a principled way to model tri-relationship among publishers, news pieces, and users simultaneously;
	\item We propose a novel framework TriFN,  which exploits both user-news interactions and publisher-news relations for learning news feature representations to predict fake news; and 
	\item We conduct extensive experiments on two real-world datasets to assess the effectiveness of TriFN.
\end{itemize}

\section{Problem Statement}
Let $\mathcal{A}=\{a_1,a_2,...,a_n\}$ be the set of $n$ news pieces,  and $\mathcal{U}=\{u_1,u_2,...,u_m\}$ be the set of $m$ users on social media posting these news pieces. We denote $\mathbf{X}\in\mathbb{R}^{n\times t}$ as the bag-of-word feature matrix of news pieces, where $t$ is the dimension of vocabulary size. We use $\mathbf{A}\in \{0,1\}^{m\times m}$ to denote the user-user adjacency matrix, where $\mathbf{A}_{ij}=1$ indicates that user $u_i$ and $u_j$ are friends; otherwise $\mathbf{A}_{ij}=0$. We denote the user-news interaction matrix as $\mathbf{W}\in\{0,1\}^{m\times n}$, where $\mathbf{W}_{ij}=1$ indicates that user $u_i$ has shared the news piece $a_j$; otherwise $\mathbf{W}_{ij}=0$. It's worth mentioning that we focus on those user-news interactions in which users agree with the news. For example, we only consider those users who share news pieces without comments, and these users share the same alignment of viewpoints with the news items~\cite{kim2017says}. We will introduce more details in Section~\ref{sec:news_user}. We also denote $\mathcal{P}=\{p_1,p_2,...,p_l\}$ as the set of $l$ news publishers. In addition, we denote $\mathbf{B}\in\mathbb{R}^{l\times n}$ as the publisher-news publishing matrix, and $\mathbf{B}_{kj}=1$ means news publisher $p_k$ publishes the news article $a_j$; otherwise $\mathbf{B}_{kj}=0$. We assume that the partisan bias labels of some publishers are given and available (see more details of how to collect partisan bias labels in Sec~\ref{sec:publisher_news}). We define $\mathbf{o}\in\{-1,0,1\}^{l\times 1}$ as the partisan label vectors, where -1, 0, 1 represents left-, neutral-, and right-partisan bias. 


Similar to previous research~\cite{shu2017fake,jin2016news}, we treat fake news detection problem as a binary classification problem. In other words, each news piece can be true or fake, and we use $\mathbf{y}=\{\mathbf{y}_1;\mathbf{y}_2;...;\mathbf{y}_n\}\in\mathbb{R}^{n\times 1}$ to represent the labels, and $\mathbf{y}_j=1$ means news piece $a_j$ is fake news; $\mathbf{y}_j=-1$ means true news. With the notations given above, the problem is formally defined as,
\\\\
\textit{Given news article feature matrix $\mathbf{X}$, user adjacency matrix $\mathbf{A}$, user social engagement matrix $\mathbf{W}$, publisher-news publishing matrix $\mathbf{B}$, publisher partisan label vector $\mathbf{o}$, and partial labeled news vector $\mathbf{y}_L$, we aim to predict remaining unlabeled news label vector $\mathbf{y}_U$.}


\section{A Tri-Relationship Embedding Framework}
\begin{figure}[tbp!]
\vspace{-0.2cm}
   \centering
   \includegraphics[scale=0.41]{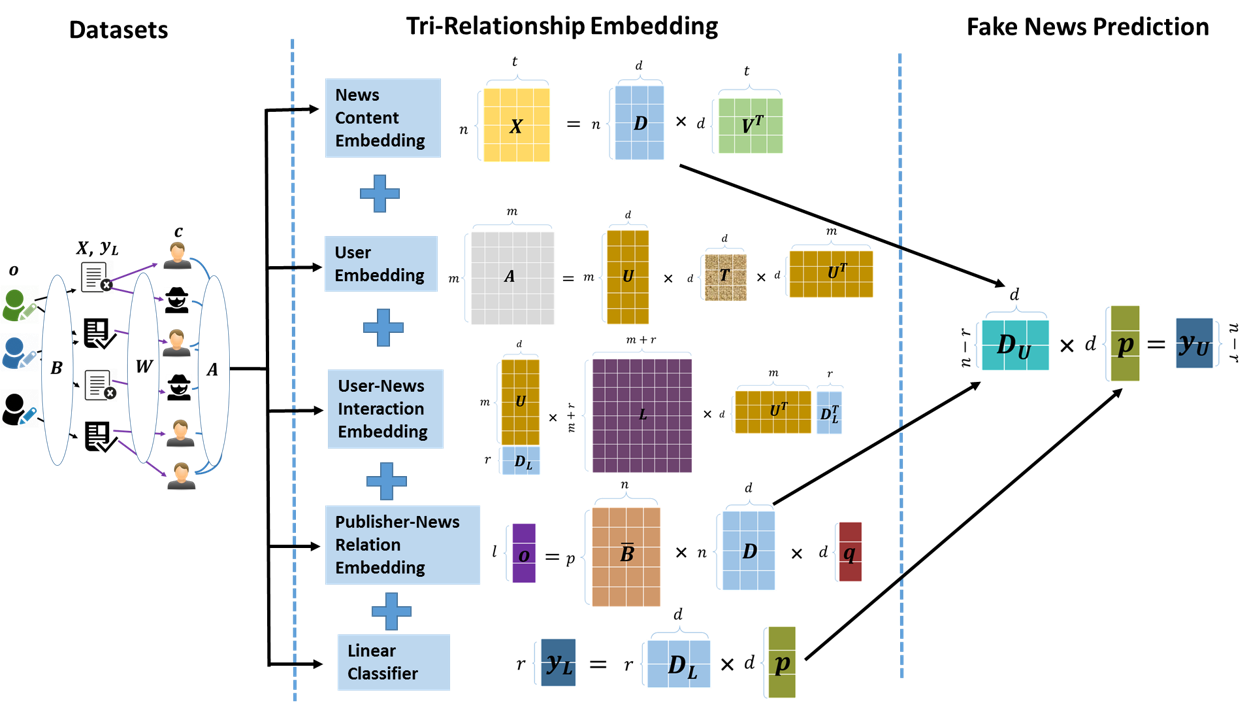}
   \vspace{0.12cm}
   \caption{The tri-relationship embedding framework, which consists of five components: news contents embedding, user embedding, user-news interaction embedding, publisher-news relation embedding, and news classification. }\vspace{-0.2cm}
   \label{fig:framework}
\end{figure}

In this section, we present the details of the proposed framework TriFN for modeling tri-relationship for fake news detection. It consists of five major components (Figure~\ref{fig:framework}): a news contents embedding component, a user embedding component, a user-news interaction embedding component, a publisher-news relation embedding component, and a semi-supervised classification component.  

In general, the news contents embedding component describes the mapping of news from bag-of-word features to latent feature space; the user embedding component illustrates the extraction of user latent features from user social relations; the user-news interaction embedding component learn the feature representations of news pieces guided by their partial labels and user credibilities; The publisher-news relation embedding component regularize the feature representations of news pieces through publisher partisan bias labels; The semi-supervised classification component learns a classification function to predict unlabeled news items. 

\subsection{News Contents Embedding}
We can use news contents to find clues to differentiate fake news and true news. Recently, it has been shown that nonnegative matrix factorization (NMF) algorithms are very practical and popular to learn document representations~\cite{xu2003document,shahnaz2006document,pauca2004text}. It can project the news-word matrix $\mathbf{X}$ to a joint latent semantic factor space with low dimensionality, such that the news-word relations are modeled as the inner product in the space. Specifically, giving the news-word matrix $\mathbf{X}\in\mathbb{R}^{n \times t}$, NMF methods try to find two nonnegative matrices $\mathbf{D}\in \mathbb{R}_{+}^{n\times d}$ and $\mathbf{V}\in \mathbb{R}_{+}^{t\times d}$, where $d$ is the dimension of the latent space, by solving the following optimization problem,

\begin{equation}
\begin{aligned}
  \min_{\substack{\mathbf{D} ,\mathbf{V}\geq0}} & \|\
  \mathbf{X} - \mathbf{D}\mathbf{V}^T\|_F^2 +\lambda (\|\mathbf{D}\|_F^2+ \|\mathbf{V}\|_F^2)\\
\end{aligned}\label{eqn:news}
\end{equation}
where $\mathbf{D}$ and $\mathbf{V}$ are the nonnegative matrices indicating low dimension representations of news pieces and words. Note that we denote $\mathbf{D}=[\mathbf{D}_L;\mathbf{D}_U]$, where $\mathbf{D}_L\in\mathbb{R}^{r\times d}$ is the news latent feature matrix for labeled news; while $\mathbf{D}_U\in\mathbb{R}^{(n-r)\times d}$ is the news latent feature matrix for unlabeled news. The term $\lambda (\|\mathbf{D}\|_F^2+ \|\mathbf{V}\|_F^2)$ is introduced to avoid over-fitting. 

\subsection{User Embedding} \label{sec:user}
On social media, people tend to form relationships with like-minded people, rather than those users who have opposing preferences and interests. Thus, connected users are more likely to share similar latent interests in news pieces. To obtain a standardized representation, we use nonnegative matrix factorization to learn the users' latent representations. Specifically, giving user-user adjacency matrix $\mathbf{A}\in\{0,1\}^{m\times m}$, we learn nonnegative matrix $\mathbf{U}\in\mathbb{R}_{+}^{m\times d}$ by solving the following optimization problem,
\begin{equation}
\begin{aligned}
  \min_{\substack{\mathbf{U},\mathbf{T}\geq 0}} & \|\mathbf{Y}\odot(\mathbf{A} - \mathbf{U} \mathbf{T}\mathbf{U}^T)\|_F^2+\lambda(\|\mathbf{U}\|_F^2+\|\mathbf{T}\|_F^2)
\end{aligned}
\end{equation}
where $\mathbf{U}$ is the user latent matrix, $\mathbf{T}\in\mathbb{R}_{+}^{d \times d}$ is the user-user correlation matrix, $\mathbf{Y}\in\mathbb{R}^{m \times m}$ controls the contribution of $\mathbf{A}$, and $\odot$ denotes the Hadamard product operation.  Since only positive links are observed in $\mathbf{A}$, following common strategies~\cite{pan2009mind}, we first set $\mathbf{Y}_{ij}=1$ if $\mathbf{A}_{ij}=1$, and then perform negative sampling and generate the same number of unobserved links and set weights as 0. The term $\lambda(\|\mathbf{U}\|_F^2+\|\mathbf{T}\|_F^2)$ is to avoid over-fitting.

\subsection{User-News Interaction Embedding} \label{sec:news_user}

We model the user-news interactions by considering the relationships between user features and the labels of news items. We have shown (see Section~\ref{sec:introduction}) that users with low credibilities are more likely to spread fake news, while users with high credibilities are less likely to spread fake news. To measure user credibility scores, we adopt the practical approach in~\cite{abbasi2013measuring}. The basic idea in~\cite{abbasi2013measuring} is that less credible users are more likely to coordinate with each other and form big clusters, while more credible users are likely to from small clusters. Specifically, the credibility scores are measured through the following major steps: 1) detect and cluster coordinate users based on user similarities; 2) weight each cluster based on the cluster size. Note that for our fake news detection task, we do not assume that credibility scores are directly provided,  but inferred from widely available data, such as user-generated contents. By using the method in~\cite{abbasi2013measuring}, we can assign each user $u_i$ a credibility score $\mathbf{c}_i\in[0,1]$. A larger $\mathbf{c}_i$ indicates that user $u_i$ has a higher credibility, while a lower $\mathbf{c}_i$ indicates a lower credibility score. We use $\mathbf{c} =\{\mathbf{c}_1, \mathbf{c}_2,...,\mathbf{c}_m\}$ to denote the credibility score vector for all users. 

First, high-credibility users are more likely to share true news pieces, so we ensure that the distance between latent features of high-credibility users and that of true news is minimized, 

\begin{equation}
\begin{aligned}
\min_{\substack{\mathbf{U},\mathbf{D_L}\geq 0}} ~\sum_{i=1}^m \sum_{j=1}^r \mathbf{W}_{ij} \mathbf{c}_i  (1 - \frac{1+\mathbf{y}_{Lj}}{2}) ||\mathbf{U}_i - \mathbf{D}_{L_j}||_2^2 
\end{aligned} \label{eqn:1}
\end{equation}
and  $(1 - \frac{1+\mathbf{y}_{Lj}}{2})$ is to ensure we only include true news pieces (i.e., $\mathbf{y}_{Lj}=-1$), and $\mathbf{c}_i$ is to adjust the contribution of user $u_i$ to the loss function. For example, if $\mathbf{c}_i$ is large (high-credibility) and $\mathbf{W}_{ij}=1$, we put a bigger weight on forcing the distance of feature $\mathbf{U}_i$ and $\mathbf{D}_{Lj}$ to be small; if $\mathbf{c}_i$ is small (low-credibility) and $\mathbf{W}_{ij}=1$, than  we put a smaller weight on forcing the distance of feature $\mathbf{U}_i$ and $\mathbf{D}_{Lj}$ to be small. 

Second, low-credibility users are more likely to share fake news pieces, and we aim to minimize the distance between latent features of low-credibility users and that of fake news,

\begin{equation}
\begin{aligned}
\min_{\substack{\mathbf{U},\mathbf{D_L}\geq 0}}~\sum_{i=1}^m \sum_{j=1}^r  \mathbf{W}_{ij} (1-\mathbf{c}_i) (\frac{1+\mathbf{y}_{Lj}}{2}) ||\mathbf{U}_i - \mathbf{D}_{L_j}||_2^2
\end{aligned} \label{eqn:2}
\end{equation}
and the term $(\frac{1+\mathbf{y}_{Lj}}{2})$ is to ensure we only include fake news pieces (i.e., $\mathbf{y}_{L_j}=1$), and $(1-\mathbf{c}_i)$ is to adjust the contribution of user $u_i$ to the loss function. For example, if $\mathbf{c}_i$ is large (high-credibility) and $\mathbf{W}_{ij}=1$, we put a smaller weight on forcing the distance of feature $\mathbf{U}_i$ and $\mathbf{D}_{Lj}$ to be small; if $\mathbf{c}_i$ is small (low-credibility) and $\mathbf{W}_{ij}=1$, then  we put a bigger weight on forcing the distance of feature $\mathbf{U}_i$ and $\mathbf{D}_{Lj}$ to be small. 

Finally, We combine Eqn~\ref{eqn:1} and Eqn~\ref{eqn:2} to consider the above two situations, and obtain the following objective function,
\begin{equation}
\begin{aligned}
\min_{\substack{\mathbf{U},\mathbf{D_L}\geq 0}}~\underbrace{\sum_{i=1}^m \sum_{j=1}^r \mathbf{W}_{ij} \mathbf{c}_i  (1 - \frac{1+\mathbf{y}_{Lj}}{2}) ||\mathbf{U}_i - \mathbf{D}_{L_j}||_2^2 }_{\text{True news}}\\+\underbrace{\sum_{i=1}^m \sum_{j=1}^r  \mathbf{W}_{ij} (1-\mathbf{c}_i) (\frac{1+\mathbf{y}_{Lj}}{2}) ||\mathbf{U}_i - \mathbf{D}_{L_j}||_2^2}_{\text{Fake news}}
\end{aligned} \label{eqn_user_engage}
\end{equation}
For simplicity, Eqn~\ref{eqn_user_engage} can be rewritten as,
\begin{equation}
\begin{aligned}
\min_{\substack{\mathbf{U},\mathbf{D_L}\geq 0}}~\sum_{i=1}^m \sum_{j=1}^r \mathbf{G}_{ij}||\mathbf{U}_i - \mathbf{D}_{L_j}||_2^2
\end{aligned}\label{eqn_user_engage1}
\end{equation}
where $\mathbf{G}_{ij} = \mathbf{W}_{ij}(\mathbf{c}_i(1-{\frac{1+\mathbf{y}_{Lj}}{2}})+(1-\mathbf{c}_i){(\frac{1+\mathbf{y}_{Lj}}{2})})$. If we denote a new matrix $\mathbf{H}=[\mathbf{U};\mathbf{D}_L]\in\mathbb{R}^{(m+r) \times d}$, we can also rewrite Eqn.~\ref{eqn_user_engage1} as a matrix form as follows,
\begin{equation}
\begin{aligned}
&\min_{\substack{\mathbf{U},\mathbf{D_L}\geq 0}}~\sum_{i=1}^m \sum_{j=1}^r \mathbf{G}_{ij}||\mathbf{U}_i - \mathbf{D}_{L_j}||_2^2
\Leftrightarrow\min_{\substack{\mathbf{H}\geq 0}}~\sum_{i=1}^m \sum_{j=1+m}^{r+m} \mathbf{G}_{ij}||\mathbf{H}_i - \mathbf{H}_{j}||_2^2
\\&\Leftrightarrow \min_{\substack{\mathbf{H}\geq 0}}~\sum_{i,j=1}^{m+r} \mathbf{F}_{ij}||\mathbf{H}_i - \mathbf{H}_{j}||_2^2 \Leftrightarrow \min_{\substack{\mathbf{H}\geq 0}}~\text{tr}(\mathbf{H}^T\mathbf{L}\mathbf{H})
\end{aligned}
\end{equation}
where $\mathbf{L}=\mathbf{S}-\mathbf{F}$ is the Laplacian matrix and $\mathbf{S}$ is a diagonal matrix with diagonal element $S_{ii}=\sum_{j=1}^{m+r}\mathbf{F}_{ij}$. $\mathbf{F}\in\mathbb{R}^{(m+r)\times(m+r)}$ is computed as follows,
\begin{equation}
\mathbf{F}_{ij} = \begin{cases}
0, & i,j \in [1,m] \text{~or~} i,j \in [m+1,m+r] \\
G_{i(j-m)}, & i\in[1,m],j\in[m+1,m+r]\\
G_{(i-m)j}, & i \in[m+1,m+r],j\in[1,m]
\end{cases}
\end{equation}

\subsection{Publisher-News Relation Embedding}\label{sec:publisher_news}
Fake news is often written to convey opinions or claims that support the partisan bias of news publishers. Thus, a good news representation should be good at predicting the partisan bias of its publisher. We obtain the list of publishers' partisan scores from a well-known media bias fact-checking websites MBFC~\footnote{https://mediabiasfactcheck.com/}. The partisan bias labels are checked with a principled methodology that ensures the reliability and objectivity of the partisan annotations. The labels are categorized into five categories: ``left'', ``left-Center'',``least-biased'',``right-Center'' and ``right''. To further ensure the accuracy of the labels, we only consider those news publishers with the annotations [``left'',``least-biased'', ``Right''], and rewrite the corresponding labels as [-1,0,1]. Thus, we can construct a partisan label vectors for news publishers as $\mathbf{o}$. Note that we may not obtain the partisan labels for all publishers, so we introduce $\mathbf{e}\in\{0,1\}^{l\times1}$ to control the weight of $\mathbf{o}$. If we have the partisan bias label of publisher $p_k$, then $\mathbf{e}_k=1$; otherwise, $\mathbf{e}_k=0$. The basic idea is to utilize publisher partisan labels vector $\mathbf{o}\in \mathbb{R}^{l\times 1}$ and publisher-news matrix $\mathbf{B}\in\mathbb{R}^{l \times n}$ to optimize the news feature representation learning. Specifically, we optimization following objective function,

\begin{equation}
\begin{aligned}
  \min_{\mathbf{D}\geq0, \mathbf{q}}~ \|\
  \mathbf{e}\odot(\bar{\mathbf{B}}\mathbf{D}\mathbf{q}-\mathbf{o})\|_2^2+\lambda\|\mathbf{q}\|_2^2
\end{aligned}
\end{equation}
where we assume that the latent feature of news publisher can be represented by the features of all the news it published, i.e., $\mathbf{\bar{\mathbf{B}}D}$. $\bar{\mathbf{B}}$ is the normalized user-news publishing relation matrix, i.e., $\bar{\mathbf{B}}_{kj} = \frac{\mathbf{B}_{kj}}{\sum_{j=1}^n\mathbf{B}_{kj}}$. $\mathbf{q}\in \mathbb{R}^{d\times 1}$ is the weighting matrix that maps news publishers' latent features to corresponding partisan label vector $\mathbf{o}$. 

\subsection{Proposed Framework - TriFN}
We have introduced how we can learn news latent features by modeling different aspects of the tri-relationship. We further employ a semi-supervised linear classifier term as follows,
\begin{equation}
\begin{aligned}
  \min_{\mathbf{p}} ~\|\
  \mathbf{D}_L \mathbf{p}-\mathbf{y}_L\|_2^2+\lambda\|\mathbf{p}\|_2^2
\end{aligned}
\end{equation}\label{eqn:prediction}
where $\mathbf{p}\in\mathbb{R}^{d\times 1}$ is the weighting matrix that maps news latent features to fake news labels. With all previous components, TriFN solves the following optimization problem,
\begin{equation}
\begin{aligned}
  \min_{\mathbf{D},\mathbf{U},\mathbf{V},\mathbf{T}\geq0, \mathbf{p},\mathbf{q}}&\|\mathbf{X} - \mathbf{D}\mathbf{V}^T\|_F^2+ \alpha\|\mathbf{Y}\odot(\mathbf{A} - \mathbf{U} \mathbf{T}\mathbf{U}^T)\|_F^2
  \\&+\beta\text{tr}(\mathbf{H}^T\mathbf{L}\mathbf{H})+
 \gamma\|\mathbf{e}\odot(\bar{\mathbf{B}}\mathbf{D}\mathbf{q}-\mathbf{o})\|_2^2
  \\&+\eta\|\mathbf{D}_L \mathbf{p}-\mathbf{y}_L\|_2^2+\lambda R
\end{aligned}\label{eqn:obj}
\end{equation}
where $R=(\|\mathbf{D}\|_F^2+\|\mathbf{V}\|_F^2+\|\mathbf{U}\|_F^2+\|\mathbf{T}\|_F^2+\|\mathbf{p}\|_2^2+\|\mathbf{q}\|_2^2)$ is to avoid over-fitting. The first term models the news latent features from news contents; the second term extracts user latent features from their social relationships; and the third term incorporates the user-news interactions; and the fourth term models publisher-news relationships. The fifth term adds a semi-supervised fake news classifier. Therefore, this framework provides a principled way to model tri-relationship for fake news prediction.

\section{An Optimization Algorithm}
In this section, we present the detail optimization process for the proposed framework TriFN. If we update the variables jointly, the objective function in Eq.~\ref{eqn:obj} is not convex. Thus, we propose to use alternating least squares to update the variables separately. For simplicity, we user $\mathcal{L}$ to denote the objective function in Eq.~\ref{eqn:obj}. Next, we introduce the updating rules for each variable in details.

\textbf{Update $\mathbf{D}$.} Let $\Psi_{D}$ be the Lagrange multiplier for constraint $\mathbf{D}\geq0$, the Lagrange  function related to $\mathbf{D}$ is,
\begin{equation}
\begin{aligned}
  \min_{\mathbf{D}} &\|\mathbf{X} - \mathbf{D}\mathbf{V}^T\|_F^2+ \beta\text{tr}(\mathbf{H}^T\mathbf{L}\mathbf{H})+\gamma\|\mathbf{e}\odot(\bar{\mathbf{B}}\mathbf{D}\mathbf{q}-\mathbf{o})\|_2^2
  \\&+\eta\|\mathbf{D}_L \mathbf{p}-\mathbf{y}_L\|_2^2+\lambda\|\mathbf{D}\|_F^2-\text{tr}(\Psi_{D}\mathbf{D}^T)
\end{aligned}
\end{equation}
and $\mathbf{D}=[\mathbf{D}_L;\mathbf{D}_U]$ and $\mathbf{H}=[\mathbf{U};\mathbf{D}_L]$. We rewrite $\mathbf{L}=[\mathbf{L}_{11},\mathbf{L}_{12};\mathbf{L}_{21},\mathbf{L}_{22}]$, where $\mathbf{L}_{11}\in\mathbb{R}^{m\times m}$, $\mathbf{L}_{12}\in\mathbb{R}^{m\times r}$,$\mathbf{L}_{21}\in\mathbb{R}^{r\times m}$, and $\mathbf{L}_{22}\in\mathbb{R}^{r\times r}$; and $\mathbf{X}=[\mathbf{X}_L,\mathbf{X}_U]$. The partial derivative of $\mathcal{L}$ w.r.t. $\mathbf{D}$ as follows,

\begin{equation}
\begin{aligned}
  \frac{1}{2}\frac{\partial \mathcal{L}}{\partial \mathbf{D}} & = (\mathbf{D}\mathbf{V}^T-\mathbf{X})\mathbf{V}+\lambda\mathbf{D}
  +\gamma\bar{\mathbf{B}}^T\mathbf{E}^T(\mathbf{E}\bar{\mathbf{B}}\mathbf{D}\mathbf{q}-\mathbf{E}\mathbf{o})\mathbf{q}^T
  \\+&\bigl[\beta\mathbf{L}_{21}\mathbf{U}+\beta\mathbf{L}_{22}\mathbf{D}_L+\eta(\mathbf{D}_L\mathbf{p}-\mathbf{y}_L)\mathbf{p}^T;\mathbf{0}\bigr]-\Psi_{D}
   \label{gradient_D}
\end{aligned}
\end{equation}
where $\mathbf{E}\in\mathbb{R}^{l\times l}$ is a diagonal matrix with $\{\mathbf{e}_k\}_{k=1}^l$ on the diagonal and zeros everywhere else.  By setting the derivative to zero and using Karush-KuhnTucker complementary condition~\cite{boyd2004convex}, i.e., $\Psi_{D}(i,j)\mathbf{D}_{ij}=0$,we get,
\begin{equation}
\begin{aligned}
	\mathbf{D}_{ij}\leftarrow\mathbf{D}_{ij}\sqrt{\frac{\hat{\mathbf{D}}(i,j)}{\tilde{\mathbf{D}}(i,j)}}
   \label{update_D}
\end{aligned}
\end{equation}
\begin{equation}
\begin{aligned}
	\hat{\mathbf{D}} &= \mathbf{X}\mathbf{V}+\gamma\bigl({\bar{\mathbf{B}}^T\mathbf{E}^T\mathbf{E}\mathbf{o}\mathbf{q}^T}\bigr)^{+}+\gamma\bigl({\bar{\mathbf{B}}^T\mathbf{E}^T\mathbf{E}\bar{\mathbf{B}}\mathbf{D}\mathbf{q}\mathbf{q}^T}\bigr)^{-}
	\\&+\bigl[\eta\bigl(\mathbf{D}_L\mathbf{p}\mathbf{p}^T\bigr)^{-}+\eta\bigl(\mathbf{y}_L\mathbf{p}^T\bigr)^{+}+\beta(\mathbf{L}_{21}\mathbf{U})^{-}
    +\beta(\mathbf{L}_{22}\mathbf{D}_L)^{-};\mathbf{0}\bigr]\\
	\tilde{\mathbf{D}} &= \mathbf{D}\mathbf{V}^T\mathbf{V}+\lambda\mathbf{D}+\gamma\bigl({\mathbf{B}}^T\mathbf{E}^T\mathbf{E}\bar{\mathbf{B}}\mathbf{D}\mathbf{q}\mathbf{q}^T)^{+}+\gamma\bigl({\bar{\mathbf{B}}^T\mathbf{E}^T\mathbf{E}\mathbf{o}\mathbf{q}^T}\bigr)^{-}
	\\&+\bigl[\beta(\mathbf{L}_{21}\mathbf{U})^{+}+\beta(\mathbf{L}_{22}\mathbf{D}_L)^{+}
	+\eta\bigl(\mathbf{D}_L\mathbf{p}\mathbf{p}^T\bigr)^{+}
  +\eta\bigl(\mathbf{y}_L\mathbf{p}^T\bigr)^{-};\mathbf{0}\bigr]
\end{aligned}
\end{equation}
where for any matrix $\mathbf{X}$, $(\mathbf{X})^+$ and $(\mathbf{X})^-$ denote the positive and negative parts of $\mathbf{X}$, respectively. Specifically, we have $(\mathbf{X})^+=\frac{ABS(\mathbf{X})+\mathbf{X}}{2}$ and $(\mathbf{X})^-=\frac{ABS(\mathbf{X})-\mathbf{X}}{2}$, $ABS(\mathbf{X})$ is the matrix with the absolute value of elements in $\mathbf{X}$.

\textbf{Update $\mathbf{U}$, $\mathbf{V}$ and $\mathbf{T}$.} The partial derivative of the Lagrange objective function w.r.t. $\mathbf{U}$ and updating rule are as follows,
\begin{equation}
\begin{aligned}
  \frac{1}{2}\frac{\partial \mathcal{L}}{\partial \mathbf{U}} & = \alpha(\mathbf{Y}\odot(\mathbf{U}\mathbf{T}\mathbf{U}^T-\mathbf{A}))\mathbf{U}\mathbf{T}^T
  +\alpha(\mathbf{Y}\odot(\mathbf{U}\mathbf{T}\mathbf{U}^T-\mathbf{A}))^T\mathbf{U}\mathbf{T}
  \\&+\lambda\mathbf{U}-\Psi_{U}+\beta(\mathbf{L}_{11}\mathbf{U}+\mathbf{L}_{12}\mathbf{D}_L)
   \label{gradient_U}
\end{aligned}
\end{equation}
\begin{equation}
\begin{aligned}
  \mathbf{U}_{ij}&\leftarrow\mathbf{U}_{ij}\sqrt{\frac{\bigl[\hat{\mathbf{U}}\bigr](i,j)}{\bigl[\tilde{\mathbf{U}}\bigr](i,j)}}
   \label{update_U}
\end{aligned}
\end{equation}
\begin{equation}
\begin{aligned}
  \hat{\mathbf{U}}& = \alpha(\mathbf{Y}\odot\mathbf{A})\mathbf{U}\mathbf{T}^T+\alpha(\mathbf{Y}\odot\mathbf{A})^T\mathbf{U}\mathbf{T}+\beta(\mathbf{L}_{11}\mathbf{U})^{-}+\beta(\mathbf{L}_{12}\mathbf{D}_L)^{-}
	\\\tilde{\mathbf{U}}&=\alpha(\mathbf{Y}\odot\mathbf{U}\mathbf{T}\mathbf{U}^T)\mathbf{U}\mathbf{T}^T+\alpha(\mathbf{Y}\odot\mathbf{U}\mathbf{T}\mathbf{U}^T)^T\mathbf{U}\mathbf{T}+\lambda\mathbf{U}
	\\&+\beta(\mathbf{L}_{11}\mathbf{U})^{+}+\beta(\mathbf{L}_{12}\mathbf{D}_L)^{+}
   \label{update_U}
\end{aligned}
\end{equation}
The partial derivatives of the Lagrange objective w.r.t $\mathbf{V}$ and updating rule are,
\begin{equation}
\begin{aligned}
  \frac{1}{2}\frac{\partial \mathcal{L}}{\partial \mathbf{V}} & = (\mathbf{D}\mathbf{V}^T-\mathbf{X})^T\mathbf{D}+\lambda\mathbf{V}-\Psi_{V}
   \label{gradient_V}
\end{aligned}
\end{equation}
\begin{equation}
\begin{aligned}
	\mathbf{V}_{ij}\leftarrow\mathbf{V}_{ij}\sqrt{\frac{\bigl[\mathbf{X}^T\mathbf{D}\bigr](i,j)}{\bigl[\mathbf{V}\mathbf{D}^T\mathbf{D}+\lambda\mathbf{V}\bigr](i,j)}}
   \label{update_V}
\end{aligned}
\end{equation}
The partial derivative of the Lagrange objective w.r.t $\mathbf{T}$ and the updating rule are,
\begin{equation}
\begin{aligned}
  \frac{1}{2}\frac{\partial \mathcal{L}}{\partial \mathbf{T}} & = \alpha \mathbf{U}^T(\mathbf{Y}\odot(\mathbf{U}\mathbf{T}\mathbf{U}^T-\mathbf{A}))\mathbf{U}+\lambda\mathbf{T}-\Psi_{T}
   \label{gradient_T}
\end{aligned}
\end{equation}
\begin{equation}
\begin{aligned}
	\mathbf{T}_{ij}\leftarrow\mathbf{T}_{ij}\sqrt{\frac{\bigl[\alpha\mathbf{U}^T(\mathbf{Y}\odot\mathbf{A})\mathbf{U}\bigr](i,j)}{\bigl[\alpha\mathbf{U}^T(\mathbf{Y}\odot\mathbf{U}\mathbf{T}\mathbf{U}^T)\mathbf{U}+\lambda\mathbf{T}\bigr](i,j)}}
   \label{update_T}
\end{aligned}
\end{equation}
\textbf{Update $\mathbf{p}$ and $\mathbf{q}$.} Optimization w.r.t $\mathbf{p}$ and $\mathbf{q}$ are essentially least square problems. By setting $\frac{\partial \mathcal{L}}{\partial \mathbf{p}}=0$ and $\frac{\partial \mathcal{L}}{\partial \mathbf{q}}=0$, the closed from solutions of $\mathbf{p}$ and $\mathbf{q}$ are as follows,
\begin{equation}
\begin{aligned}
	\mathbf{p}&=(\eta \mathbf{D}_L^T\mathbf{D}_L+\lambda \mathbf{I})^{-1}\eta\mathbf{D}_L^T \mathbf{y}_L\\
	\mathbf{q}&=(\gamma \mathbf{D}^T\bar{\mathbf{B}}^T\mathbf{E}\bar{\mathbf{B}}\mathbf{D}+\lambda\mathbf{I})^{-1}\gamma \mathbf{D}^T\bar{\mathbf{B}}^T\mathbf{E}\mathbf{o}
   \label{update_PQ}
\end{aligned}
\end{equation}
Where $\mathbf{I}$ is an identity matrix, and $\mathbf{E}\in\mathbb{R}^{l\times l}$ with $\mathbf{e}_k,k=1,\dots,l$ on the diagonal and zeros everywhere else.  

\subsection{Optimization Algorithm of TriFN}
We present the details to optimize TriFN in Algorithm~\ref{alg_sfn}. We first randomly initialize $\mathbf{U}, \mathbf{V}, \mathbf{T},\mathbf{D}, \mathbf{p}, \mathbf{q}$ in line~\ref{alg_ini}, and construct the Laplacian matrix $\mathbf{L}$ in line~\ref{alg_lap}. Then we repeatedly update related parameters through Line~\ref{alg_D} to Line~\ref{alg_PQ} until convergence. Finally, we predict the labels of unlabeled news $\mathbf{y}_U$ in line~\ref{alg_pre}. The convergence of Algorithm~\ref{alg_sfn} is guaranteed because the objective function is nonnegative and in each iteration it will monotonically decrease the objective value, and finally it will converge to an optimal point~\cite{lee2001algorithms}.

\begin{algorithm}[tb!]
\caption{The optimization process of TriFN framework}
\label{alg:sfn}
\begin{algorithmic}[1]
\Require $\mathbf{X}, \mathbf{A}, \mathbf{B}, \mathbf{W}, \mathbf{Y},\mathbf{o}, \mathbf{y}_L,\alpha,\beta,\gamma,\lambda, \eta$ 
\Ensure $\mathbf{y}_U$
\State Randomly initialize $\mathbf{U}, \mathbf{V}, \mathbf{T},\mathbf{D}, \mathbf{p}, \mathbf{q}$ \label{alg_ini}
\State Precompute Laplacian matrix $\mathbf{L}$ \label{alg_lap}
\Repeat
\State Update $\mathbf{D}$ with Eqn~\ref{update_D} \label{alg_D}
\State Update $\mathbf{U}$ with Eqn~\ref{update_U}
\State Update $\mathbf{V}$ with Eqn~\ref{update_V}
\State Update $\mathbf{T}$ with Eqn~\ref{update_T} 
\State Update $\mathbf{p}$,$\mathbf{q}$ with Eqn~\ref{update_PQ} \label{alg_PQ}
\Until{convergence}
\State Calculate $\mathbf{y}_U = \text{Sign}(\mathbf{D}_U\mathbf{p})$ \label{alg_pre}
\end{algorithmic}\label{alg_sfn}
\end{algorithm}

\subsection{Complexity Analysis}

The main computation cost comes from the fine-tuning variables for Algorithm~\ref{alg_sfn}. In each iteration, the time complexity for computing $\mathbf{D}$ is $\mathcal{O}(nd+nld^2+rd+rm+n^2)$. Similarly, the computation cost for $\mathbf{V}$ is approximately $\mathcal{O}(tnd)$, for $\mathbf{U}$ is $\mathcal{O}(m^4d^3+md)$, for $\mathbf{T}$ is about $\mathcal{O}(m^4d^3+m^2d^2)$. To update $\mathbf{p}$ and $\mathbf{q}$, the costs are approximately $\mathcal{O}(d^3+d^2+dr)$ and $\mathcal{O}(d^2ln+d^3+dl)$. The overall time complexity is the sum of the costs of initialization and fine-tuning.

\section{Experiments}
In this section, we present the experiments to evaluate the effectiveness of the proposed TriFN framework. Specifically, we aim to answer the following research questions:
\begin{itemize}
\item Is TriFN able to improve fake news classification performance by modeling publisher partisan and user engagements simultaneously?
\item How effective are publisher partisan bias modeling and user engagement learning, respectively, in improving the fake news detection performance of TriFN?
\item Can the proposed method handle early fake news detection when limited user engagements are provided?
\end{itemize}



\subsection{Datasets}

\begin{table}[htb!]
\small
\centering \caption{The statistics of FakeNewsNet dataset}
\begin{tabular}{l|cc}
\toprule
 Platform &BuzzFeed & PolitiFact  \\
\midrule
\# Users & 15,257& 23,865\\
\midrule
\# Engagements & 25,240 & 37,259 \\
\midrule
\# Social Links & 634,750& 574,744\\
\midrule
 \# Candidate news & 182& 240\\
\midrule
\# True news & 91& 120\\
\midrule
\# Fake news & 91 & 120\\
\midrule
\# Publisher & 9& 91\\
\bottomrule
\end{tabular} \label{tab:data}
\end{table}
\vspace{-0.2cm}
We utilize one of the comprehensive fake news detection benchmark dataset called FakeNewsNet~\cite{shu2017fake,shu2018fakenewsnet}. The dataset is collected from two platforms with fact-checking: \textit{BuzzFeed} and \textit{PolitiFact}, both containing news content with labels and social context information. News content includes the meta attributes of the news (e.g., body text), and social context includes the related user social engagements of news items (e.g., user posting/sharing news in Twitter). The detailed statistics of the datasets are shown in Table~\ref{tab:data}.

\subsection{Experimental Settings}
To evaluate the performance of fake news detection algorithms, we use the following metrics, which are commonly used to evaluate classifiers in related areas: Accuracy, Precision, Recall, and F1.
We randomly choose 80\% of news pieces for training and remaining 20\% for testing, and the process is performed for 10 times and the average performance is reported. We compare the proposed framework TriFN with several state-of-the-art fake news detection methods. Existing methods mainly focus on extracting\textit{discriminative features} and feed them into a classification algorithm to differentiate fake news. Next, we introduce several representative features as follows,

\begin{itemize}
\item \textbf{RST}~\cite{rubin2015towards}: RST stands for Rhetorical Structure Theory, which builds a tree structure to represent rhetorical relations among the words in the text. RST can extract style-based features of news by mapping the frequencies of rhetorical relations to a vector space~\footnote{The code is available at: https://github.com/jiyfeng/DPLP}. 
\item \textbf{LIWC}~\cite{pennebaker2015development}: LIWC stands for Linguistic Inquiry and Word Count, which is widely used to extract the lexicons falling into psycholinguistic categories. It's based on a large sets of words that represent psycholinguistic processes, summary categories, and part-of-speech categories. It learns a feature vector from a psychology and deception perspective~\footnote{The readers can find more details about the software and feature description at: http://liwc.wpengine.com/}.
\item \textbf{Castillo}~\cite{castillo2011information}: Castillo extract various kinds of features from those users who have shared a news item on social media. The features are extracted from user profiles and friendship network. We also include the credibility score of users inferred in Sec~\ref{sec:news_user} as an additional social context feature. 
\item \textbf{RST+Castillo}: RST+Castillo represents the concatenated features of RST and  Castillo, which include features extracted from both news content and social context. 
\item \textbf{LIWC+Castillo}: LIWC+Castillo represents the concatenated features of LIWC and Castillo, which consists of feature information from both news content and social context. 
\end{itemize}

Note that for a fair and comprehensive comparison, we choose the above feature extraction methods from following aspects: 1) only extract features from \textbf{news contents}, such as RST, LIWC; 2) only construct features from \textbf{social context}, such as Castillo; and 3) consider both \textbf{news content and social context}, such as RST+Castillo, LIWC+Castillo.

\begin{table*} [htbp!]
\vspace{-0.15cm}
\centering \caption{Best performance comparison for fake news detection}
\begin{tabular}{|l|l|c|c|c|c|c|c|}
\hline
Datasets & Metric  &RST & LIWC & Castillo & RST+Castillo & LIWC+Castillo & TriFN\\
\hline \hline
\multirow{4}{*}{\textbf{BuzzFeed}}   & Accuracy  & $0.600\pm0.063$ & $0.719\pm0.074$ & $0.800\pm0.037$ & $0.816\pm0.052$ & $0.825\pm0.052$  & $\textbf{0.864}\pm\textbf{0.026}$ \\
\cline{2-8}
&Precision &$0.662\pm0.109$ & $0.722\pm0.077$ & $0.822\pm0.077$ & $0.879\pm0.049$ & $0.821\pm0.061$ & $\textbf{0.849}\pm\textbf{0.040}$ \\
\cline{2-8}

&Recall &$0.615\pm0.018$ & $0.732\pm0.171$ & $0.776\pm0.027$ & $0.748\pm0.098$ & $0.829\pm0.055$  & $\textbf{0.893}\pm\textbf{0.013}$ \\
\cline{2-8}
&F1 &$0.633\pm0.056$ & $0.709\pm0.075$ & $0.797\pm0.044$ & $ 0.805\pm0.066$& $0.822\pm0.035$ & $\textbf{0.870}\pm\textbf{0.019}$ \\
\hline
\hline
\multirow{4}{*}{\textbf{PolitiFact}} & Accuracy &  $0.604\pm0.060$  & $0.688\pm0.063$ & $0.796\pm0.052$ & $0.838\pm0.036$ & $0.829\pm0.052$  & $\textbf{0.878}\pm\textbf{0.017}$ \\
\cline{2-8}
&Precision &$0.564\pm0.064$ & $0.725\pm0.087$ & $0.767\pm0.056$  & $0.851\pm0.052$ & $0.821\pm0.116$ & $\textbf{0.867}\pm\textbf{0.034}$ \\
\cline{2-8}
&Recall &$0.705\pm0.148$  & $0.617\pm0.100$ & $0.889\pm0.044$ & $0.824\pm0.063$ &  $0.879\pm0.047$ & $\textbf{0.893}\pm\textbf{0.023}$ \\
\cline{2-8}
&F1 &$0.615\pm0.074$ & $0.666\pm0.092$ & $0.822\pm0.037$ & $0.835\pm0.043$ & $0.843\pm0.054$ & $\textbf{0.880}\pm\textbf{0.015}$ \\
\hline
\end{tabular} \label{tab:performance}
\end{table*}
\subsection{Performance Comparison}
We evaluate the effectiveness of the proposed framework TriFN for fake news classification. We determine model parameters with cross-validation strategy, and we repeat the generating process of training/test set for three times and the average performance is reported. We first perform cross validation on parameters $\lambda \in \{0.001,0.01,0.1,1,10\}$, and choose those parameters that achieves best performance, i.e., $\lambda=0.1$. We also choose latent dimension $d=10$ for easy parameter tuning, and focus on the parameters that contribute the tri-relationship modeling components. The parameters for TriFN are set as $\{\alpha=1e-4, \beta=1e-5, \gamma=1,\eta=1\}$ for BuzzFeed and $\{\alpha=1e-5, \beta=1e-4, \gamma=10,\eta=1\}$ for PolitiFact.

\begin{table} [htbp!]
\small
\centering \caption{Average F1 of baselines for different learning algorithms on BuzzFeed. Best scores are highlighted.}
\begin{tabular}{lccccc}
\toprule
  \thead{Method}&\thead{RST} & \thead{LIWC} & \thead{Castillo} & \thead{RST \\+Castillo} & \thead{LIWC \\+Castillo} \\
\midrule
 LogReg  & $0.519$ & $0.660$ & $0.714$ & $0.728$ & $0.760$\\
\hline 
NBayes &$0.511$ & $0.370$ & $0.600$ & $0.716$ & $0.680$\\
\hline 
DTree &$0.566$ & $0.581$ & $0.736$ & $0.681$ & $0.772$ \\
\hline 
RForest &$0.538$ & $\cellcolor{gray!75}0.709$ & $0.767$ & $ \cellcolor{gray!75}0.805$& $0.733$\\
\hline 
XGBoost &$0.480$ & $0.672$ & $\cellcolor{gray!75}0.797$  & $0.795$ & $0.782$ \\
\hline 
AdaBoost &$\cellcolor{gray!75}0.633$ & $0.701$ & $0.724$ & $0.791$ &  $0.768$\\
\hline 
GradBoost &$0.492$ & $0.699$ & $0.772$ & $0.724$ &  $\cellcolor{gray!75}0.822$\\
\hline
\end{tabular} \label{tab:avg_buzz}
\end{table}

\begin{table}[htbp!]
\small
\centering \caption{Average F1 of baselines for different learning algorithms on PolitiFact. Best scores are highlighted.}
\begin{tabular}{lccccc}
\toprule
  \thead{Method}&\thead{RST} & \thead{LIWC} & \thead{Castillo} & \thead{RST \\+Castillo} & \thead{LIWC \\+Castillo} \\
\midrule
 LogReg  & $\cellcolor{gray!75}0.615$ & $0.432$ & $0.707$ & $0.668$ & $0.653$\\
\hline 
NBayes &$0.537$ & $0.486$ & $0.442$ & $0.746$ & $0.687$\\
\hline 
DTree &$0.514$ & $0.661$ & $0.771$ & $0.792$ & $0.772$ \\
\hline 
RForest &$0.463$ & $0.586$ & $0.767$ & $ \cellcolor{gray!75}0.835$& $0.836$\\
\hline 
XGBoost &$0.552$ & $0.648$ & $\cellcolor{gray!75}0.822$  & $0.783$ & $0.823$ \\
\hline 
AdaBoost &$0.502$ & $\cellcolor{gray!75}0.666$ & $0.800$ & $0.787$ &  $0.831$\\
\hline 
GradBoost &$0.517$ & $0.650$ & $0.818$ & $0.803$ &  $\cellcolor{gray!75}0.843$\\
\hline
\end{tabular} \label{tab:avg_politi}
\end{table}
We test the baseline features on different learning algorithms, and choose the one that achieves the best performance (see Table~\ref{tab:performance}). The algorithms include Logistic Regression (LogReg for short), Na\"ive Bayes (NBayes), Decision Tree (DTree), Random Forest (RForest), XGBoost, AdaBoost, and Gradient Boosting (GradBoost). We used the open-sourced \textit{xgboost}~\cite{chen2016xgboost} package and \textit{scikit-learn}~\cite{pedregosa2011scikit} machine learning framework in Python to implement all these algorithms. To ensure a fair comparison of features, we ran all the algorithms using default parameter settings. We also show the performances for each learning algorithm and report the average performance on both datasets. Due to the space limitation, we only show the results of F1 score (Table~\ref{tab:avg_buzz} and Table~\ref{tab:avg_politi}). We observe similar results for other metrics in terms of average performance. Based on Table~\ref{tab:performance}, Table~\ref{tab:avg_buzz}, and Table~\ref{tab:avg_politi}, we have following observations:

\begin{itemize}
\item For news content based methods RST and LIWC, we can see that $LIWC>RST$ for both best performance and average performance, indicating that LIWC can better capture the linguistic features in news contents. The good results of LIWC demonstrate that fake news pieces are very different from real news in terms of choosing the words that reveal psychometrics characteristics. 
\item In addition, social context based features are more effective than news content based features, i.e., $Castillo>RST$ and $Castillo>LIWC$. It shows that social context features have more discriminative power than those only on news content for predicting fake news.
\item Moreover, methods using both news contents and social context perform better than those methods purely based on news contents, and those methods only based on social engagements, i.e., $LIWC+Castillo>LIWC\text{~or~}Castillo$ and $RST+Castillo>RST\text{~or~}Castillo$. This indicates that features extracted from news content and corresponding social context have complementary information, and thus boost the detection performance.
\item Generally, for methods based on both news content and social context (i.e., RST+Castillo, LIWC+Castillo, and TriFN), we can see that TriFN consistently outperforms the other two baselines, i.e., $TriFN>LIWC+Castillo$ and $TriFN>RST+Castillo$, in terms of all evaluation metrics on both datasets. For example, TriFN achieves average relative improvement of $4.72\%, 5.84\%$ on BuzzFeed and $5.91\%, 4.39\%$ on PolitiFact, comparing with LIWC+Castillo in terms of $Accuracy$ and $F1$ score. It supports the importance to model tri-relationship of publisher-news and news-user to better predict fake news.

\end{itemize}
\subsection{Assessing Impacts of Users and Publishers}
In previous section, we observe that TriFN framework improves the classification results significantly. In addition to news contents, we also captures user-news interactions and publisher-news relations. Now, we investigate the effects of these components by defining three variants of TriFN:
\begin{itemize}
\item TriFN\textbackslash P - We eliminate the effect of publisher partisan modeling part $\gamma\|\mathbf{e}\odot(\bar{\mathbf{B}}\mathbf{D}\mathbf{q}-\mathbf{o})\|_2^2$ by setting $\gamma=0$.
\item TriFN\textbackslash S - We eliminate the effects of user social engagements components $\alpha\|\mathbf{Y}\odot(\mathbf{A} - \mathbf{U} \mathbf{T}\mathbf{U}^T)\|_F^2+\beta\text{tr}(\mathbf{H}^T\mathbf{L}\mathbf{H})$ by setting $\alpha, \beta=0$.
\item TriFN\textbackslash PS - We eliminate the effects of both publisher partisan and user social engagements, by setting $\alpha, \beta, \gamma=0$. The model only consider news content embedding.
\end{itemize}
\begin{figure}[htbp!]
\centering
\subfigure[BuzzFeed]{
  {\includegraphics[scale=0.45]{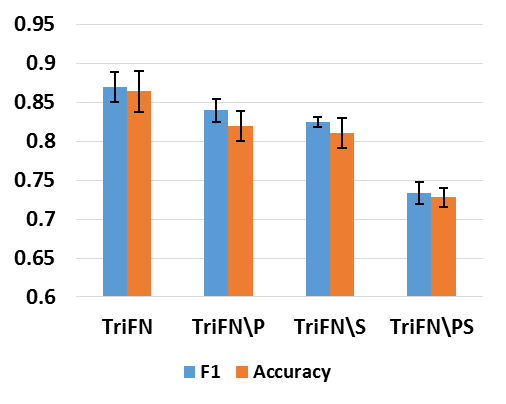}}
}
~~
\subfigure[PolitiFact]{
  {\includegraphics[scale=0.45]{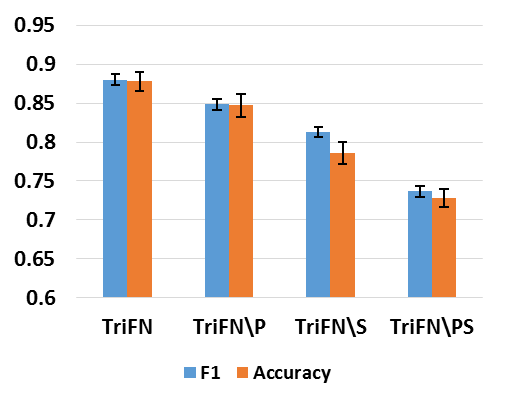}}
}
\caption{Impact analysis of users and publishers for fake news detection.}\label{fig:component}
\end{figure}

The parameters in all the variants are determined with cross-validation and the best performances are shown in Figure~\ref{fig:component}, we have following observations:
\begin{itemize}
  \item When we eliminate the effect of user social engagements component , the performance of TriFN\textbackslash S degrades in comparison with TriFN. For example, the performance reduces $5.2\%$ and $6.1\%$ in terms of F1 and Accuracy metrics on BuzzFeed, $7.6\%$ and $10.6\%$ on PolitiFact. The results suggest that social engagements in TriFN is important.
  \item We have similar observations for TriFN\textbackslash P when eliminating the effect of publisher partisan component. The results suggest the importance to consider publisher-news relations through publisher partisan bias in TriFN.
  \item When we eliminate both components in TriFN\textbackslash PS, the results are further reduced compared to TriFN\textbackslash S and TriFN\textbackslash P. It also suggests that components of user-news and publisher-news embedding are complementary to each other.
\end{itemize}
Through the component analysis of TriFN, we conclude that (i) both components can contribute to the performance improvement of TriFN; (ii) it's necessary to model both news contents and social engagements because they contain complementary information.

\subsection{Early Fake News Detection}

\begin{figure}[htb!]
\vspace{-0.2cm}
\centering
\subfigure[Accuracy on BuzzFeed]{
  {\includegraphics[scale=0.27]{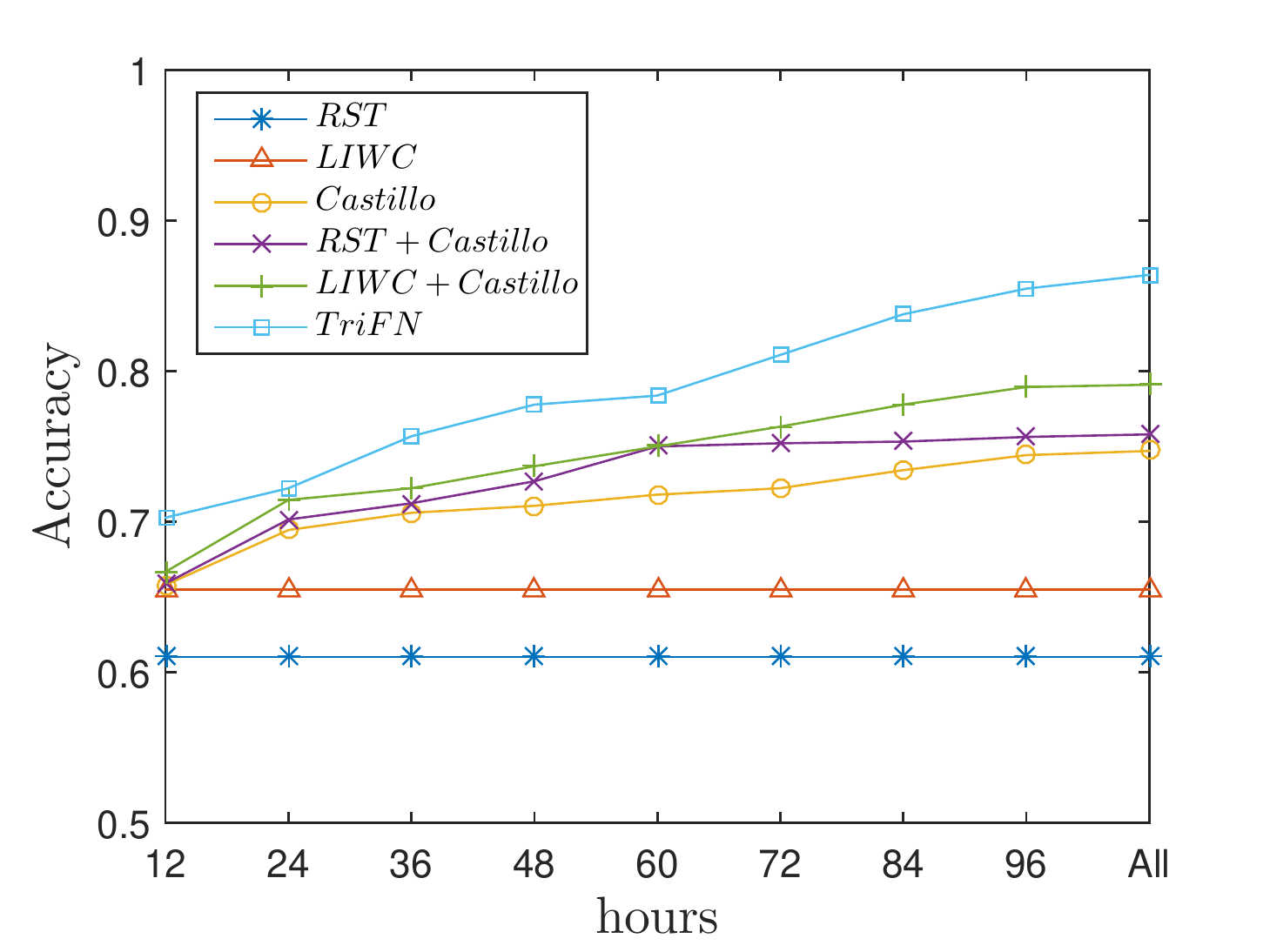}}
}
\subfigure[F1 on BuzzFeed]{
  {\includegraphics[scale=0.27]{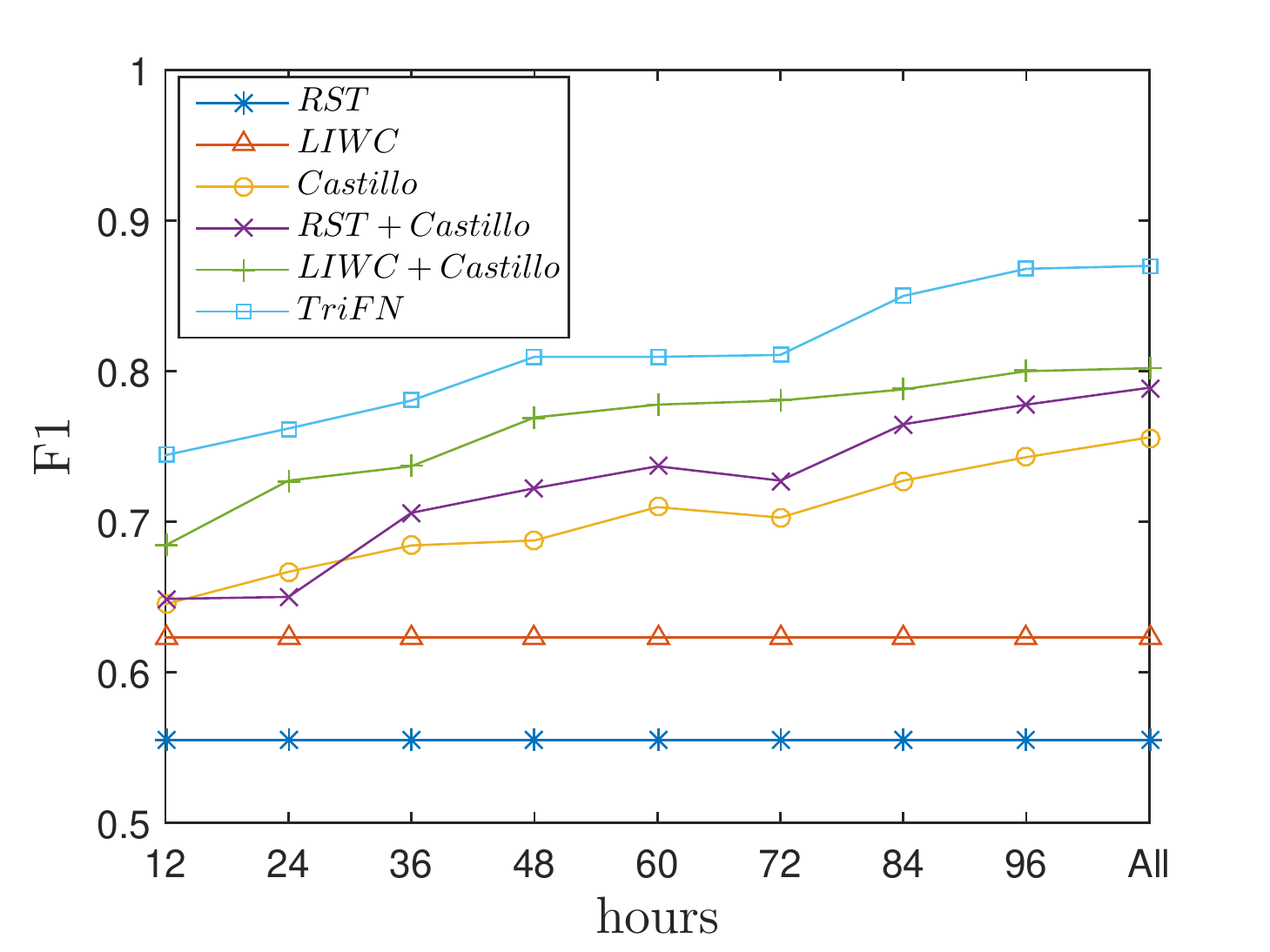}}
}
\subfigure[Accuracy on PolitiFact]{
  {\includegraphics[scale=0.27]{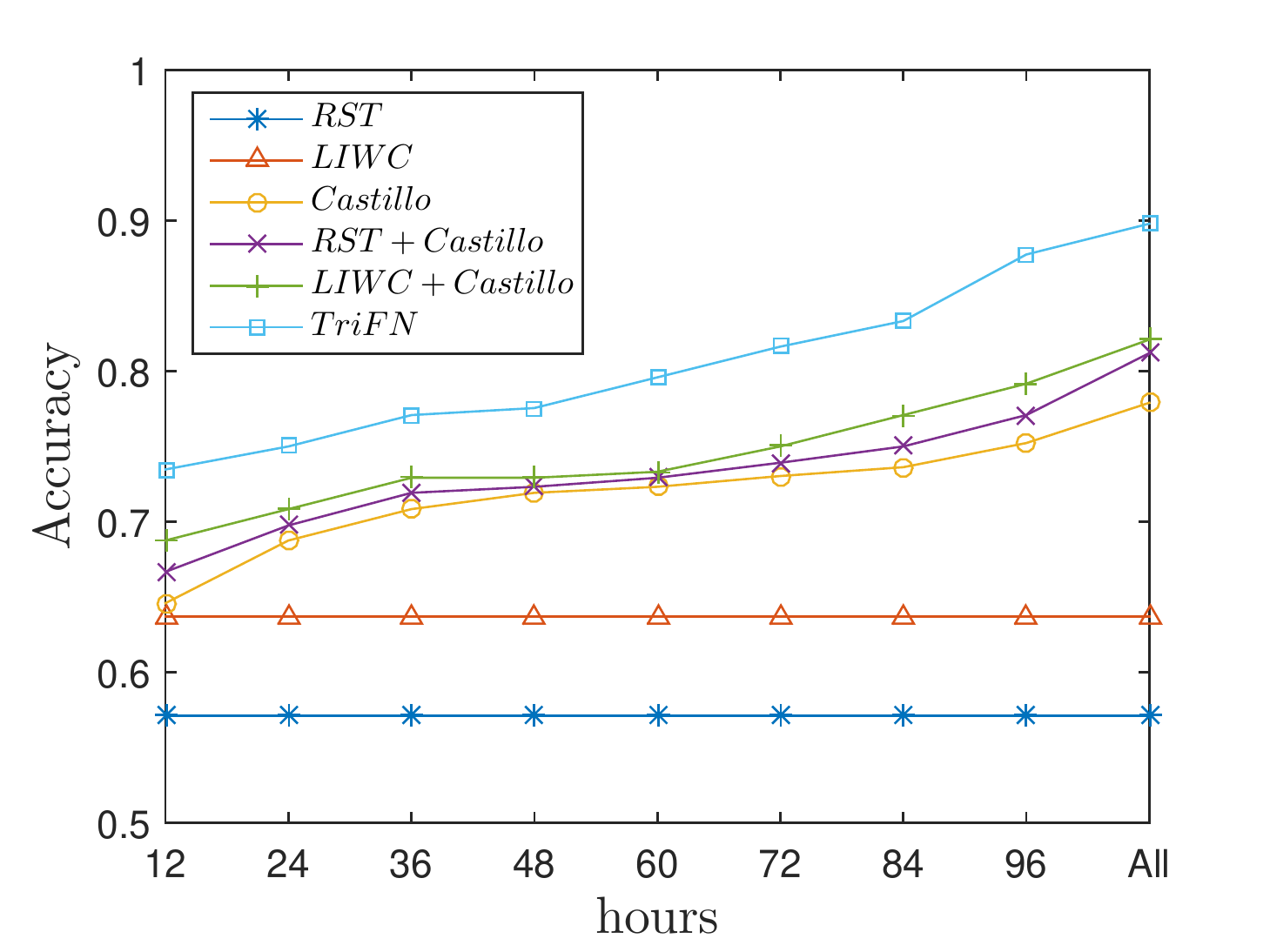}}
}
\subfigure[F1 on PolitiFact]{
  {\includegraphics[scale=0.27]{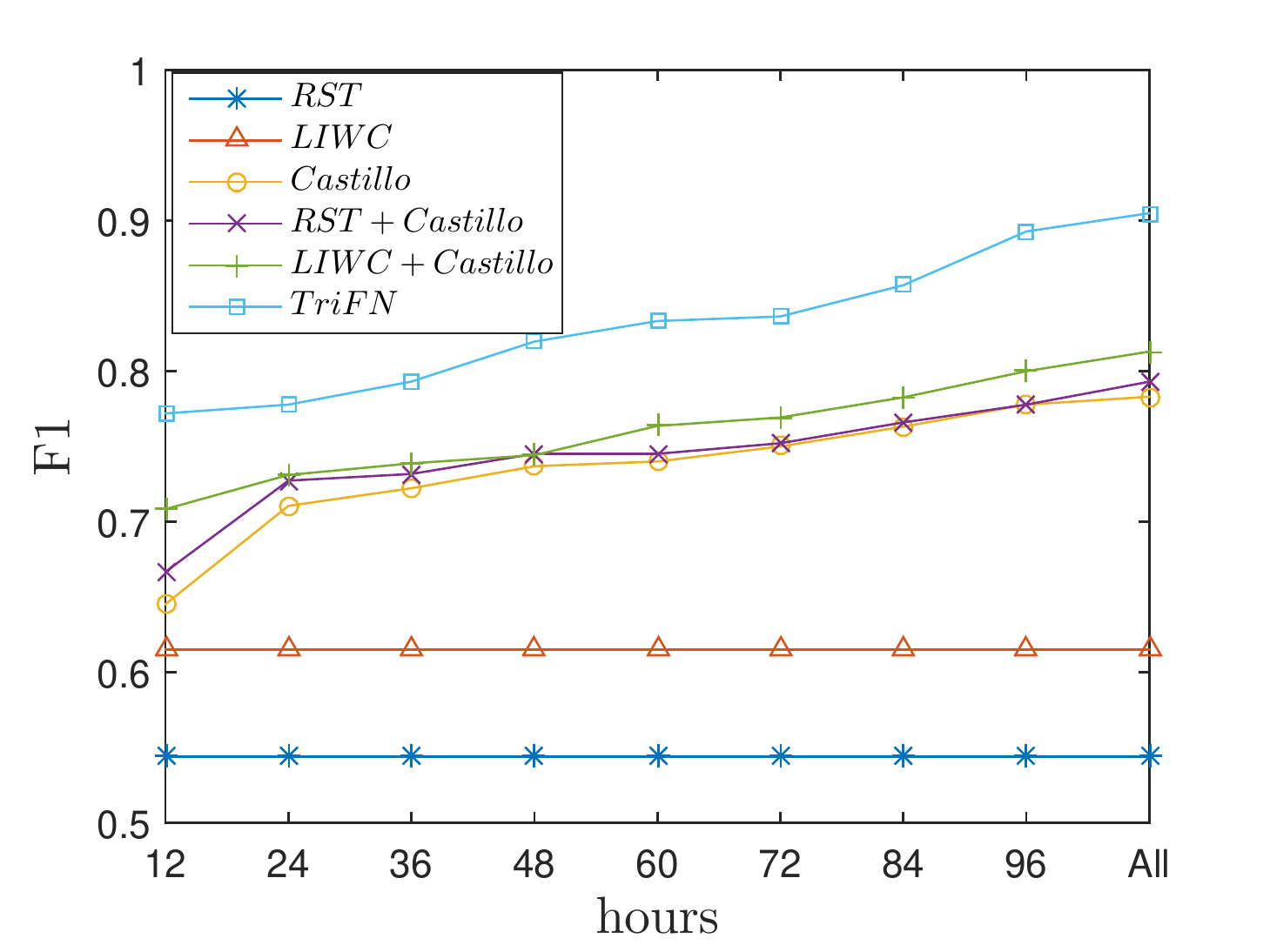}}
}
\caption{The performance of early fake news detection on BuzzFeed and PolitiFact in terms of Accuracy and F1. }\label{fig:early}
\end{figure}

Early detection of fake news is very desirable to restrict the dissemination scope of fake news and prevent the future propagation on social media. Early fake news detection aims to give early alert of fake news, by only considering limited social context within a specific range of time delay of original news posted. Specifically, we change the delay time in $[12,24,36,48,60,72,84,96]$ hours. 
From Figure~\ref{fig:early}, we can see that: 1) generally, the detection performance is getting better when the delay time increase for those methods using social context information, which indicates that more social engagements of users on social media provide more additional information for fake news detection; 2) The proposed TriFN consistently achieves best performances on both datasets for accuracy and F1, which demonstrate the importance of embedding user-news interactions to capture effective feature representations; and 3) Even in the very early stage after fake news has been published, TriFN can already achieve good performance. For example, TriFN can achieve F1 score more than 80\% within 48 hours on both datasets, which shows promising potentials to combat fake new at the early stage.

\vspace{-0.2cm}

\subsection{Model Parameter Analysis}\label{sec:para} 
\begin{figure}[!htbp]
\vspace{-0.2cm}
\centering
\subfigure[$\eta$ and $\gamma$ on BuzzFeed]{
  {\includegraphics[scale=0.25]{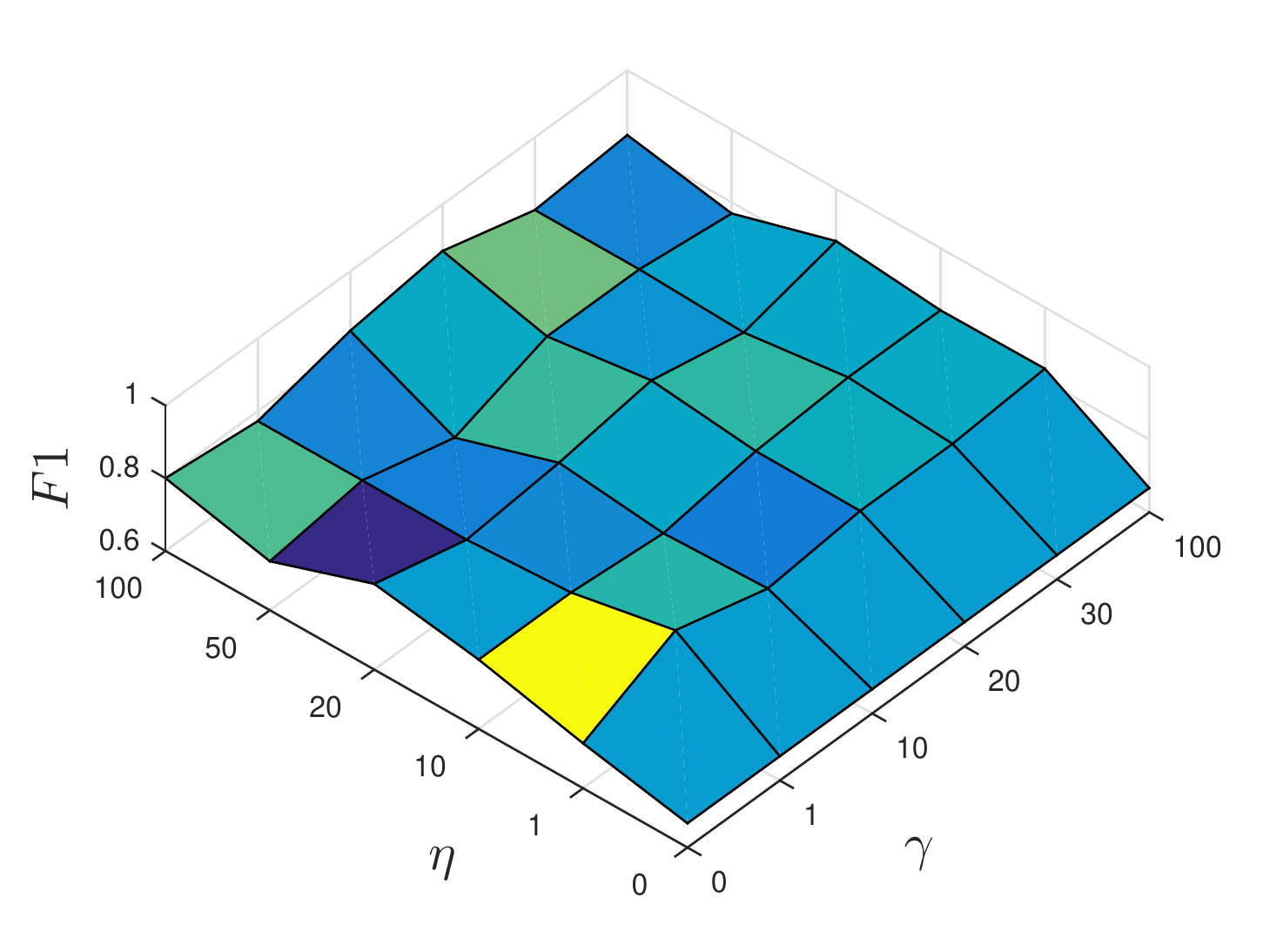}}
}
\subfigure[$\eta$ and $\gamma$ on PolitiFact]{
  {\includegraphics[scale=0.25]{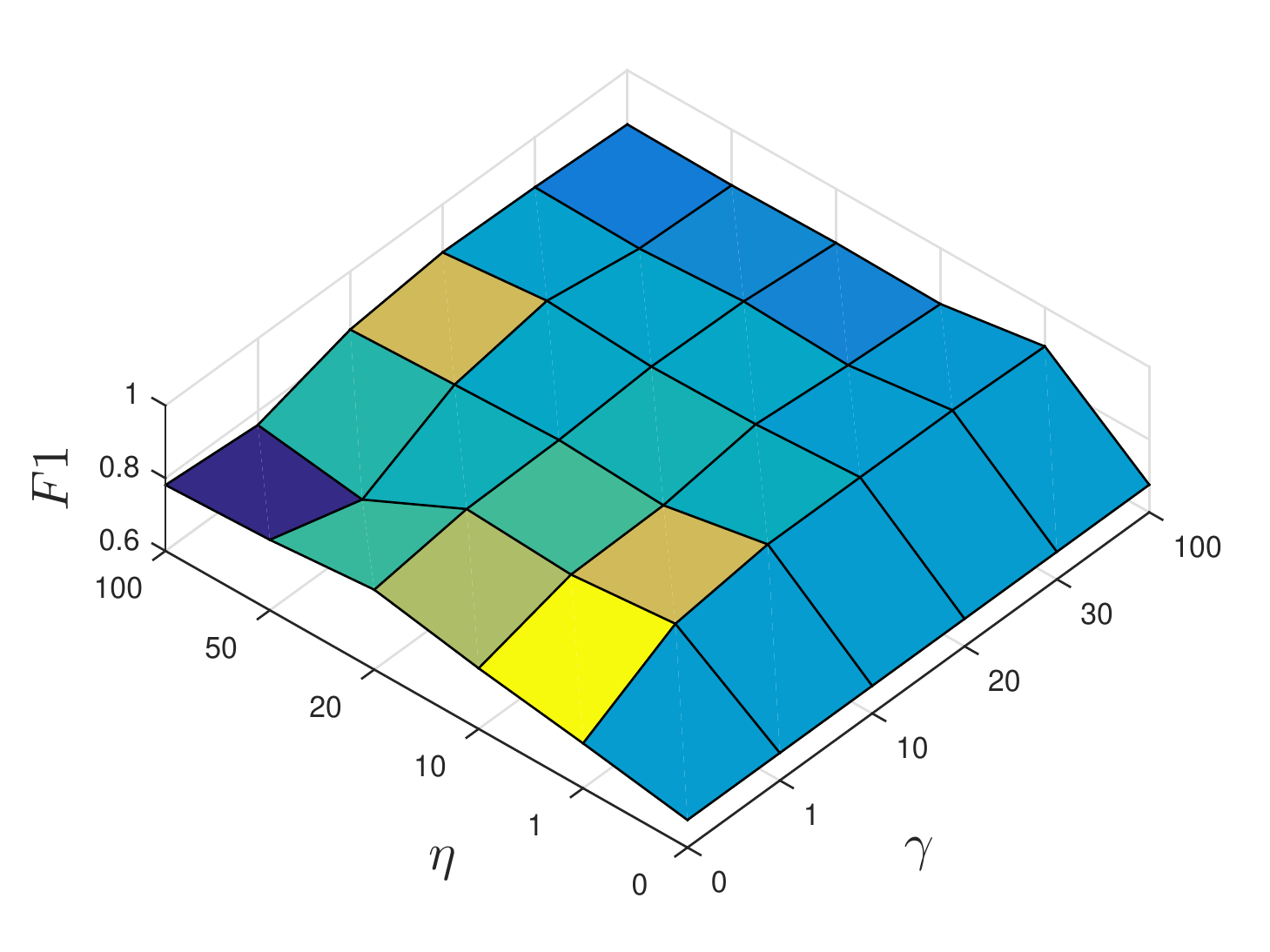}}
}
\vspace{-0.15cm}
\subfigure[$\alpha$ and $\beta$ on BuzzFeed]{
  {\includegraphics[scale=0.25]{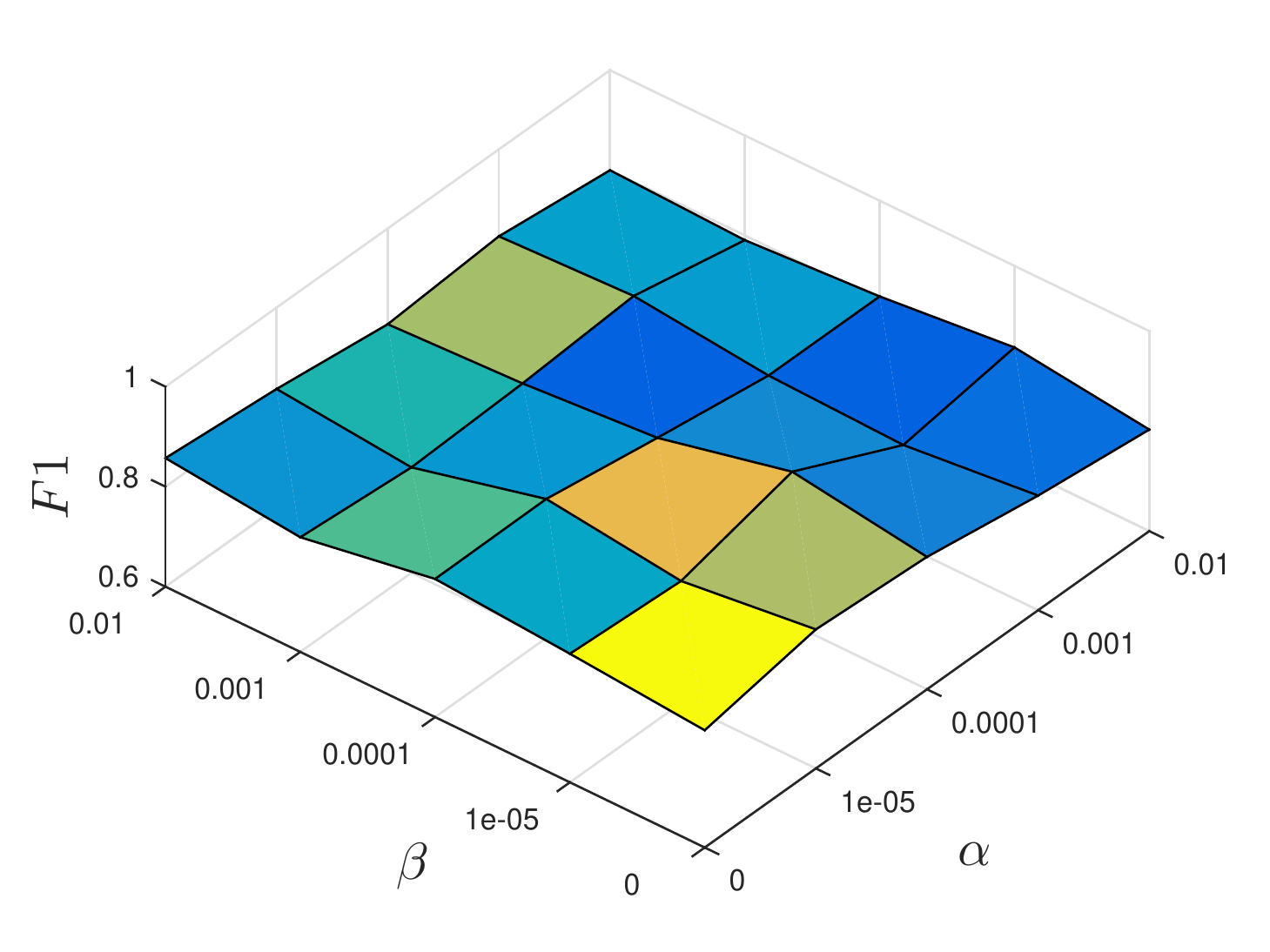}}
}
\subfigure[$\alpha$ and $\beta$ on PolitiFact]{
  {\includegraphics[scale=0.25]{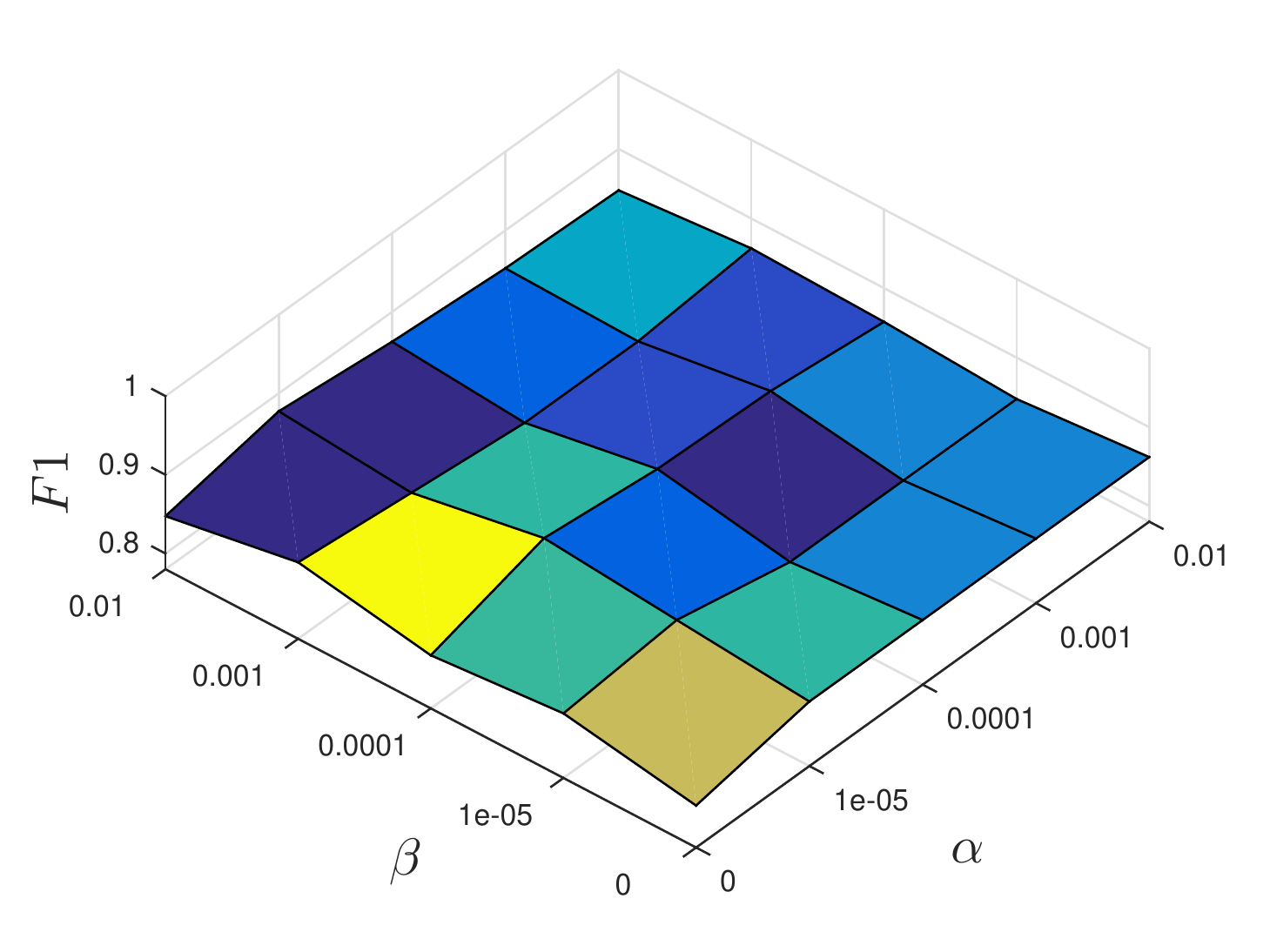}}
}
\caption{Model parameter analysis for TriFN on BuzzFeed and PolitiFact in terms of F1.}\label{fig:gamma_eta}
\end{figure}

The proposed TriFN has four important parameters. The first two are $\alpha$ and $\beta$, which control the contributions from social relationship and user-news engagements. $\gamma$ controls the contribution of publisher partisan and $\eta$ controls the contribution of semi-supervised classifier. We first fix $\{\alpha=1e-4, \beta=1e-5\}$ and $\{\alpha=1e-5, \beta=1e-4\}$ for BuzzFeed and PolitiFact, respectively. Then we vary $\eta$ as $\{1,10,20,50,100\}$ and $\gamma$ in $\{1,10,20,30,100\}$. The performance variations are depicted in Figure~\ref{fig:gamma_eta}. We can see i) when $\eta$ increases from 0, eliminating the impact of semi-supervised classification term, to 1, the performance increase dramatically in both datasets. These results support the importance to combine semi-supervised classifier to feature learning; ii) generally, the increase of $\gamma$ will increase the performance in a certain region, $\gamma\in[1,50]$ and $\eta\in[1,50]$ for both datasets, which easy the process for parameter setting. Next, we fix $\{\gamma=1, \eta=1\}$ and $\{\gamma=10, \eta=1\}$ for BuzzFeed and PolitiFact, respectively. Then we vary $\alpha, \beta \in[0,1e-5,1e-4,1e-3,0.001,0.01]$. We can see that i) when $\alpha$ and $\beta$ increase from 0, which eliminate the social engagements, to $1e-5$, the performance increases relatively, which again support the importance of social engagements; ii) The performance tends to increase first and then decrease, and it's relatively stable in $[1e-5,1e-3]$.

\section{Related Work}\label{sec:related}

We briefly introduce the related work about fake news detection on social media. Fake news detection methods generally focus on using \textit{news contents} and \textit{social contexts}~\cite{shu2017fake,zhou2018tutorial}.

News contents contain the clues to differentiate fake and real news. For news content based approaches, features are extracted as linguistic-based and visual-based. Linguistic-based features capture specific writing styles and sensational headlines that commonly occur in fake news content~\cite{potthast2017stylometric}, such as lexical and syntactic features. Visual-based features try to identify fake images~\cite{gupta2013faking} that are intentionally created or capturing specific characteristics for images in fake news. News content based models include i) knowledge-based: using external sources to fact-checking claims in news content~\cite{magdy2010web,wu2014toward}, and 2) style-based: capturing the manipulators in writing style, such as deception~\cite{feng2012syntactic,rubin2015truth} and non-objectivity~\cite{potthast2017stylometric}. For example, Potthast \textit{et al.}~\cite{potthast2017stylometric} extracted various style features from news contents and predict fake news and media bias.

In addition to news content, social context related to news pieces contains rich information to help detect fake news. For social context based approaches, the features mainly include user-based, post-based and network-based. User-based features are extracted from user profiles to measure their characteristics and credibilities~\cite{castillo2011information,kwon2013prominent,kai2019unsupervised,shu2018understanding}. For example, Shu \textit{et al.}~\cite{shu2018understanding} proposed to understand user profiles from various aspects to differentiate fake news. Yang \textit{et al.}~\cite{kai2019unsupervised} proposed an unsupervised fake news detection algorithm by utilizing users' opinions on social media and estimating their credibilities. Post-based features represent users' social response in term of stance~\cite{jin2016news}, topics~\cite{ma2015detect}, or credibility~\cite{castillo2011information,wu2018tracing}. Network-based features~\cite{fakebookchapter} are extracted by constructing specific networks, such as diffusion network~\cite{kwon2013prominent} etc. Social context models basically include stance-based and propagation-based. Stance-based models utilize users' opinions towards the news to infer news veracity~\cite{jin2016news}. Propagation-based models assume that the credibility of news is highly related to the credibilities of relevant social media posts, which several propagation methods can be applied~\cite{jin2016news}. Recently, deep learning models are applied to learn the temporal and linguistic representation of news~\cite{shu2018fakenewstracker,wang2018eann,karimi2018multi}. Shu \textit{et al.}~\cite{shu2018clickbait} proposed to generate synthetic data for augmenting training data to help improve the detection of clickbaits. It's worth mentioning that we can not directly compare the propagation-based approaches, because we assume we only have user actions, e.g., posting the news or not. In this case, the propagation signals inferred from text are the same and thus become ineffective.

In this paper, we are to our best knowledge the first to classify fake news by learning the effective news features through the tri-relationship embedding among publishers, news contents, and social engagements.



\section{Conclusion and Future Work}\label{sec:conc}
Due to the inherent relationship among publisher, news and social engagements during news dissemination process on social media, we propose a novel framework TriFN to model tri-relationship for fake news detection. TriFN can extract effective features from news publisher and user engagements separately, as well as capture the interrelationship simultaneously. Experimental results on real world fake news datasets demonstrate the effectiveness of the proposed framework and importance of tri-relationship for fake news prediction. It's worth mentioning TriFN can achieve good detection performance in the early stage of news dissemination.

There are several interesting future directions. First, it's worth to explore effective features and models for early fake news detection, as fake news usually evolves very fast on social media; Second, how to extract features to model fake news intention from psychology's perspective needs further investigation. At last, how to identify low quality or even malicious users spreading fake news is important for fake news intervention and mitigation. 

\section{Ackowledgments}
This material is based upon work supported by, or in part by, the NSF \#1614576 and the ONR grant N00014-16-1-2257.

\balance
\bibliographystyle{ACM-Reference-Format}
\bibliography{abbrv}

\end{document}